\documentclass[twocolumn,showpacs,preprintnumbers,amsmath,amssymb]{revtex4}
\usepackage{amssymb}
\usepackage{txfonts}
\usepackage{bbm}
\usepackage{graphicx}
\usepackage{appendix}
\usepackage{epsf}

\begin{document}

\title{Spin-Density-Wave and Asymmetry of Coherence Peaks in iron-Pnictide Superconductors from a two-orbital model}

\author
{Tao Zhou, Degang Zhang, and C. S. Ting}

\affiliation{Texas Center for Superconductivity and Department of
Physics, University of Houston, Houston, Texas 77204, USA\\
}

\date{\today}

\begin{abstract} We study theoretically the coexistence of the
spin-density-wave (SDW) and superconductivity in electron-doped
iron-pnictide superconductors based on the two orbital model and
Bogoliubov-de Gennes equations. The phase diagram is mapped out and
the evolution of the Fermi surface as the
 doping varies is presented. The local density of states
 has also been calculated from low to high doping. We show that the strength
 of the superconducting coherence peak at the positive energy gets enhanced and the one at the negative energy
 is suppressed by
 the SDW order in the underdoped region. Several features of our results are in
 good agreement with the experiments.

 \end{abstract}

 \maketitle

The discovery of the iron-based superconducting (SC)
materials~\cite{kam} has attracted much attention. These materials
share some common features of the high-T$_c$ cuprates~\cite{mik},
with high transition temperature, layered system and similar phase
diagram. The parent compounds are with spin-density-wave (SDW) or
antiferromagnetic (AF) order, and they are not Mott insulators but
poor 'metals'. By doping either holes or electrons into the parent
compound, the SDW/AF order becomes weakened, and superconductivity
shows up. Theoretically the Fermi surface nesting is proposed to be
responsible for the presence of the SDW order~\cite{dong}. Moreover, the
mechanism for the superconductivity in this system is still unclear
while many suggested that the correlation effect is still essential
and the spin fluctuation should be responsible for the
superconductivity~\cite{maz}.

 A quite
intriguing question is how the magnetic (SDW/AF) order competes with
the superconductivity and whether the magnetic order could coexist
with the SC order at low doping densities. This is still an open
issue and the result is under debate or depends on the materials. It
was reported that the magnetic phase and SC phase are totally
separated without the coexisting region in the fluorine doped
material CeFeAsO$_{1-x}$F$_{x}$ according to the inelastic neutron
scattering (INS) experiments~\cite{zhao}. However, it was also
reported that the magnetic order and the SC order coexist at low
doping densities in the Co-doped material
BaFe$_{2-x}$Co$_x$As$_2$~\cite{pra,les,jul}. The coexistence of the
two orders is also reported in some other materials, such as
Ba$_{1-x}$K$_{x}$Fe$_2$As$_2$~\cite{jul,chen},
Sr$_{1-x}$K$_x$Fe$_2$As$_2$~\cite{zha}, and
SmFeAsO$_{1-x}$F$_x$~\cite{liu}. Theoretically, it was proposed that
the coexistence of these two orders is possible and can explain some
experimental observations~\cite{parker}. While relatively this
subject remains less explored on the theoretical front.

The motivation of the present work is to fill this void and examine
this issue for electron-doped samples. We adopt a two-orbital model
by taking into account two Fe ions per unit cell, as proposed in
Ref.~\cite{zhang} by one of the present authors. We map out the
phase diagram and show that the doping evolution of the Fermi
surface based on this model is consistent with the angle resolved
photoemission spectroscopy (ARPES) experiments on the electron-doped
samples BaFe$_{2-x}$Co$_x$As$_2$~\cite{tera,sek}.  We calculate the
spatially distributed magnetic order and SC order self-consistently
based on the Bogoliubov-de Gennes (BdG) equations. The SC order here
is chosen to have $s_{\pm}$-wave symmetry which is supported by some
experimental observations~\cite{naka} and theoretical
studies~\cite{maz,chub}. The obtained SDW phase in which the
electron spin is ferromagnetic along $y$-direction and
antiferromagnetic along $x$-direction is in agreement with the
experiments~\cite{cruz,che}. The magnitude of magnetic order
decreases as the doping increases. The superconductivity occurs as
the doping $\delta\geq 0.01$. The magnetic order coexists with the
SC order at low doping ($0.01\leq\delta\leq0.12$). The local density
of states (LDOS) is calculated and  the signatures of the
coexistence of the SDW and SC orders are also identified.

We start with an effective model with the hopping elements, pairing
terms, and on-site interactions, expressed by,
\begin{equation}
H=H_t+H_{\Delta}+H_{int}.
\end{equation}
The first term is the hopping term, expressed by,
\begin{equation}
H_t=-\sum_{i\mu j\nu\sigma}(t_{i\mu
j\nu}c^\dagger_{i\mu\sigma}c_{j\nu\sigma}+h.c.)-t_0\sum_{i\mu\sigma}c^{\dagger}_{i\mu\sigma}c_{i\mu\sigma},
\end{equation}
where $i,j$ are the site indices and $\mu,\nu=1,2$ are the orbital
indices. $t_0$ is the chemical potential.

$H_\Delta$ is the pairing term,
\begin{equation}
H_\Delta=\sum_{i\mu j\nu\sigma}(\Delta_{i\mu
j\nu}c^\dagger_{i\mu\sigma}c^{\dagger}_{j\nu\bar{\sigma}}+h.c.).
\end{equation}

$H_{int}$ is the on-site interaction term. Following
Ref~\cite{andr}, we here include the Coulombic interaction and Hund
coupling $J_H$. At the mean-field level, the interaction Hamiltonian
can be written as~\cite{jiang},
\begin{eqnarray}
H_{int}=&&U\sum_{i\mu\sigma\neq\bar{\sigma}} \langle
n_{i\mu\bar{\sigma}}\rangle
n_{i\mu\sigma}+U^{\prime}\sum_{i,\mu\neq\nu,\sigma\neq\bar{\sigma}}\langle
n_{i\mu\bar{\sigma}}\rangle n_{i\nu{\sigma}}
\nonumber\\
&&+(U^{\prime}-J_H)\sum_{i,\mu\neq\nu,\sigma}\langle
n_{i\mu\sigma}\rangle n_{i\nu\sigma}.
\end{eqnarray}
where $n_{i\mu\sigma}$ are the density operators at the site $i$ and
orbital $\mu$. $U^{\prime}$ is taken to be $U-2J_H$~\cite{andr}.

Then the Hamiltonian can be diagonalized by solving the BdG
equations self-consistently,
\begin{equation}
\sum_j\sum_\nu \left( \begin{array}{cc}
 H_{i\mu j\nu\sigma} & \Delta_{i\mu j\nu}  \\
 \Delta^{*}_{i\mu j\nu} & -H^{*}_{i\mu j\nu\bar{\sigma}}
\end{array}
\right) \left( \begin{array}{c}
u^{n}_{j\nu\sigma}\\v^{n}_{j\nu\bar{\sigma}}
\end{array}
\right) =E_n \left( \begin{array}{c}
u^{n}_{i\mu\sigma}\\v^{n}_{i\mu\bar{\sigma}}
\end{array}
\right),
\end{equation}
where the Hamiltonian $H_{i\mu j\nu\sigma}$ is expressed by,
\begin{eqnarray}
H_{i\mu j\nu\sigma}=&&-t_{i\mu j\nu}+[U\langle
n_{i\mu\bar{\sigma}}\rangle+(U-2J_H)\langle
n_{i\bar{\mu}\bar{\sigma}}\rangle\nonumber \\&&+(U-3J_H)\langle
n_{i\bar{\mu}\sigma}\rangle-t_0]\delta_{ij}\delta_{\mu\nu}.
\end{eqnarray}

The SC order parameter and the local electron density $\langle
n_{i\mu}\rangle$ satisfy the following self-consistent conditions,
\begin{eqnarray}
\Delta_{i\mu j\nu}=\frac{V_{i\mu j\nu}}{4}\sum_n
(u^{n}_{i\mu\uparrow}v^{n*}_{j\nu\downarrow}+u^{n}_{j\nu\uparrow}v^{n*}_{i\mu\downarrow})\tanh
(\frac{E_n}{2K_B T}),
\end{eqnarray}
\begin{eqnarray}
\langle n_{i\mu}\rangle &=&\sum_n
|u^{n}_{i\mu\uparrow}|^{2}f(E_n)+\sum_n
|v^{n}_{i\mu\downarrow}|^{2}[1-f(E_n)].
\end{eqnarray}
Here $V_{i\mu j\nu}$ is the pairing strength and $f(x)$ is the Fermi
distribution function.

The LDOS is expressed by,
\begin{equation}
\rho_{i}(\omega)=\sum_{n\mu}[|u^{n}_{i\mu\sigma}|^{2}\delta(E_n-\omega)+
|v^{n}_{i\mu\bar{\sigma}}|^{2}\delta(E_n+\omega)],
\end{equation}
where the delta function $\delta(x)$ is taken as
$\Gamma/\pi(x^2+\Gamma^2)$, with the quasiparticle damping
$\Gamma=0.004$ (the results are qualitatively the same for different
$\Gamma$). The supercell
 technique~\cite{jzhu} is used to calculate the LDOS.

We use the hopping constant suggested by Ref.~\cite{zhang}, namely,
\begin{eqnarray}
t_{i\mu,i\pm\hat{\alpha}\mu}&=&t_1 \qquad (\alpha=\hat{x},\hat{y})\\
t_{i\mu,i\pm(\hat{x}+\hat{y})\mu}&=&\frac{1+(-1)^{i}}{2}t_2+\frac{1-(-1)^{i}}{2}t_3\\
t_{i\mu,i\pm(\hat{x}-\hat{y})\mu}&=&\frac{1+(-1)^{i}}{2}t_3+\frac{1-(-1)^{i}}{2}t_2\\
t_{i\mu,i\pm\hat{x}\pm\hat{y}{\nu}}&=&t_4 \qquad (\mu\neq \nu) .
\end{eqnarray}

The pairing symmetry is determined by the pairing potential $V_{i\mu
j\nu}$. We have carried out extensive calculations to search for
favorable pairing symmetry based on the present band structure.
Especially, the next-nearest-neighbor intra-orbital potential will
produce $s_{x^2y^2}$-wave ($s_\pm$-wave) order parameter, i.e.,
$\Delta \propto \cos k_x \cos k_y/(\cos k_x+\cos k_y)$ in the
extended/reduced Brillouin zone (the Brillouin zone reduces to half
of the extended one considering two Fe ions per unit cell). Here the
pairing symmetry we obtained is independent on the initial input
values and the results is consistent with previous calculation based
on a different two-orbital model~\cite{jiang}. In the following
calculation we will focus on this pairing symmetry.

Throughout the work, we use the hopping constant $t_{1-4}=1, 0.4,
-2, 0.04$. $t_0$ is determined by the electron filling per site $n$
$(n=2+\delta)$. The on-site Coulombic interaction and Hund coupling
$U$ and $J_H$ are taken as $U=3.4$ and $J_H=1.3$, respectively,
consistent with the recent estimation~\cite{yang}. The pairing
potential is chosen as $V=1.2$. The numerical calculation is
performed on $20\times 20$ lattice with the periodic boundary
conditions. An $80\times 80$ supercell is taken to calculate the
LDOS.

\begin{figure}
\centering
  \includegraphics[width=3.in]{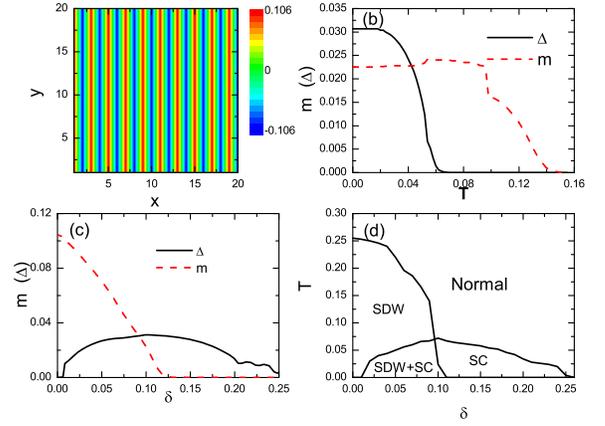}
\caption{(Color online) (a) The intensity plot of the magnetic order
at zero doping and zero temperature. (b) The magnitude of the SC
order parameter $\Delta$ and magnetic order $m$ as a function of the
temperature with the doping density $\delta=0.09$. (c)  The
magnitude of the SC order parameter $\Delta$ and magnetic order $m$
as a function of the doping at zero temperature. (d) The calculated
phase diagram. }
\end{figure}

We plot the spatial distribution of the magnetic order
[$m_i=\frac{1}{4}\sum_{\mu}(n_{i\mu\uparrow}-n_{i\mu\downarrow})$]
at the zero temperature and zero doping in Fig.1(a). As seen, the
magnetic spin order is antiferromagnetic along the $x$ direction and
ferromagnetic along the $y$ direction, corresponding to the
$(\pi,0)/(\pi,\pi)$ SDW in the extended/reduced Brillouin zone. This
result is consistent with the INS experiments~\cite{cruz,che} and
previous theoretical calculation based on different band
structure~\cite{jiang}. There exists another degenerate SDW state
with the spin order is antiferromagnetic along $y$ direction and
ferromagnetic along $x$ direction. The four-fold symmetry breaking
and the presence of the SDW order are due to the Fermi surface
feature nesting at low doping which will be discussed below. The
magnetic order decreases as the temperature increases or doping
increases. The amplitudes of the SC order and magnetic order as
functions of the temperature and doping are shown in Figs.1(b) and
1(c), respectively. At the fixed doping, both the magnetic order and
SC order decrease as the temperature increases and two transition
temperatures are revealed, as seen in Fig.1(b). The
superconductivity occurs at the doping $0.01$ and vanishes at the
doping 0.26, as seen in Fig.1(c). The magnetic order is maximum at
zero doping and decreases monotonically as the doping increases. The
calculated phase diagram is plotted in Fig.1(d). As seen, the
magnetic order and SC order coexist in the underdoped region. The
magnetic order decreases abruptly and a quantum critical point
$\delta=0.12$ is revealed. The superconductivity appears as the
magnetic order is suppressed and the SC transition temperature T$_c$
reaches the maximum as the magnetic order disappears. Our results
are reasonably consistent with the experiments on the
BaFe$_{2-x}$Co$_x$As$_2$~\cite{pra,les,jul}.

The real-space mean-field Hamiltonian can be transformed to the
momentum space because the SC order is uniform and the magnetic
order has the period $2\times 1$. The Fermi surface, defined by the
zero energy contours of the quasiparticles, can be obtained through
the momentum space Hamiltonian. The evolution of the normal state
Fermi surface with increasing doping densities is shown in Fig.2.

\begin{figure}
\centering
  \includegraphics[width=3.in]{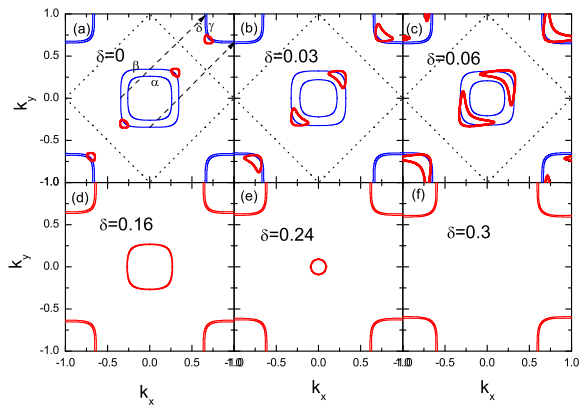}
\caption{(Color online) The doping evolution of the Fermi surface at
the temperature $T=0.1$ (i.e., the superconductivity disappears).
The (red) bold solid curves and (black) dotted lines in panels (a-c)
indicate the ungapped Fermi surfaces and the magnetic Brillouin zone
boundary in the magnetic phase, respectively. The (blue) thin solid
curves in panels (a)-(c) show the normal states Fermi surface by
setting the magnetic order $m=0$.}
\end{figure}

At zero doping [Fig.2(a)], the Fermi surface in the normal state
[(blue) thin lines] contains two hole pockets around point
$\Gamma=(0,0)$ (labeled as $\alpha$ and $\beta$ bands) and two
electron pockets around point $M=(\pi,\pi)$ (labeled as $\delta$ and
$\gamma$ bands). The existence of the SDW here represents the
pairing of the electron and hole. Thus in the SDW phase, the
momentum of the SDW order should be consistent with the Fermi
surface nesting momentum connecting the hole pockets and electron
pockets. As seen in Fig.2(a), the prime nesting vector (denoted by
the dashed arrows) in the reduced Brilliouin zone is $(\pi,\pi)$. As
the doping increases, the hole pockets shrink. Then the nesting
vector should deviate
 from $(\pi,\pi)$. But with the presence of the  underline lattice, the nesting vector is still pinned at
 $(\pi,\pi)$ and this is what has been observed by experiments~\cite{les}, thus the magnetic order
will decrease as the doping increases.
  When the doping increases to
  $\delta\thicksim0.1$ (at $T=0.1$), there is no nesting vector that
  can be sustained by the lattice and the SDW order disappears.
The magnetic Fermi surfaces [red (bold) solid curves] as shown in
Figs.2(a-c) are represented by ungapped pockets along the $\Gamma-M$
line or $k_x=k_y$ direction. Parts of the original Fermi surface
[(blue) thin solid curves] are gapped by the SDW order. The small
ungapped Fermi surface pockets along the $\Gamma-M$ line is
consistent with experiments on the parent compound
BaAs$_2$Fe$_2$~\cite{dingh}. The existence of ungapped Fermi surface
pockets at zero doping [see Fig.2(a)] indicates that the parent
compound is in fact not an 'insulator' a but a poor 'metal'. Here
the four-fold symmetry is broken in the magnetic phase, due to the
presence of the magnetic order shown in Fig.1(a). There exists
another degenerate solution with the Fermi surface pockets along the
$k_x=-k_y$ direction, corresponding to the $(0,\pi)$-SDW ground
state in the extended Brillouin zone. As the doping increases,  the
ungapped pockets become larger and larger and eventually cover the
whole Fermi surface as the SDW order disappears.

As the doping increases to $\delta=0.16$, and $0.24$ [Figs.2(d) and
2(e)], the $\alpha$-band is filled by electrons completely and the
Fermi surface of this band disappears, thus only three Fermi surface
pockets are left. This feature is consistent with the
experiments~\cite{tera}. As the doping density increases to about
$0.30$ [Fig.2(f)], the Fermi surface pocket of $\beta$-band will
also disappear and only two electron-like Fermi surface pockets are
left, which is also consistent with the experiments~\cite{sek}. On
the other hand, the electron pockets expand a little as the doping
increases but relatively the electron pockets depend weakly on the
doping, also in agreement with the ARPES
experiments~\cite{tera,sek}.

\begin{figure}
\centering
  \includegraphics[width=3.in]{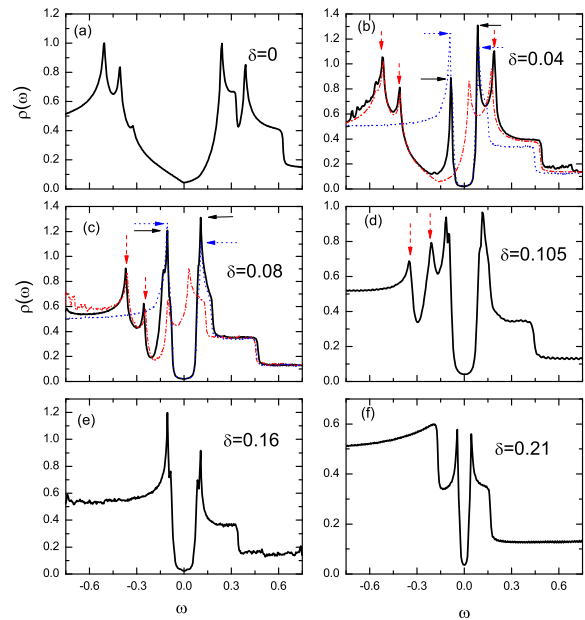}
\caption{(Color online) The LDOS spectra with different doping
densities at the zero temperature. The (red) dash-dotted and (blue)
dotted lines in panels (b) and (c) are non-selfconsistent results
with $\Delta\equiv0$, and $m\equiv0$, respectively. }
\end{figure}

We also calculate the LDOS spectra according to Eq. (9). The spectra
at different doping densities are shown in Fig.3. At zero doping
[Fig.3(a)], the density of states shows four well-defined coherence
peaks due to the SDW order. The maximum dip of the spectrum is at
the chemical potential or zero energy. Note that there are two
coherence peaks at negative energies and two at positive energies.
The splitting of the coherence peaks is caused by the inter-orbital
coupling $t_4$. As the doping increases [Figs.3(b) and 3(c)], the
maximum dip of the LDOS due to the $(\pi,\pi)$ SDW order is nearly
independent of doping and always pinned at the chemical potential of
zero doping. This is similar to the case of d-density wave in
cuprate superconductors~\cite{zhu}. As we define the zero energy at
the chemical potential, the maximum dip of the SDW spectrum will
shift to the left or the negative energy as doping increases. While
in the SC state, two SC coherence peaks show up, and the mid-gap
point is always located at the zero energy or the chemical potential
of the electron-doped system.

We now discuss the LDOS spectra in the coexisting region. In this
region, the pure SC state and the pure SDW state have higher free
energy. These two pure states can be obtained by setting the
magnetic order $m\equiv0$ and the SC order is calculated
self-consistently and setting the SC order $\Delta\equiv0$ and the
magnetic order is obtained self-consistently, respectively. The
results of the pure SC and SDW states are represented by
 (blue) dotted and (red) dash-dotted lines, respectively in Figs.3(b) and 3(c). In
the pure SDW state, four SDW peaks are obtained with the maximum dip
at the negative energy. In the pure SC state, two SC coherence peaks
are seen with the mid point of the gap at the zero energy. In the
coexisting state, the structure of the spectra inside the SC gap is
practically the same to the spectra of the pure SC state. As shown
in Figs.3(b) and 3(c), we can see clearly the previous SDW peaks
outside the SC gap. But the SDW peaks inside the SC gap in the pure
magnetic state cannot be seen in the coexisting state. The
additional peaks outside the SC peaks, positioned by the (red)
vertical arrows can be regarded as one signature of the coexistence
of the magnetic and SC orders in electron-doped materials. As doping
increases, only negative SDW peaks would appear, as seen in
Figs.3(c) and 3(d). It is needed to point out that such structures
outside the SC gap due to the SDW order as shown in Figs.3(a-d) so
far have not been clearly identified by the experiments.

Another significant feature caused by the magnetic order can be seen
from the intensities of the SC coherence peaks [see Figs.3(b) and
3(c)]. In pure SC state, the intensity of the SC peak at the
negative energy is higher than that at the positive energy for all
the doping densities we considered. These SC peaks are positioned by
the parallel (blue) dotted arrows. In the coexisting region, as we
mentioned above, the SDW spectra shift to the negative energy so
that the SC peak at the negative energy is within the SDW gap and
that at the positive energy is outside the SDW gap. Thus the
intensity of the SC peak at the positive energy is enhanced and the
one at the negative energy gets suppressed by the SDW order. As a
result, in the coexisting state, the intensity of the SC coherence
peak [denoted by the (black) solid arrows] at the negative energy is
lower than that at the positive energy. The asymmetry disappears
near the optimal doping $(\approx0.105)$, as seen in Fig.3(d). In
the overdoped region [see Figs.3(e) and 3(f)], the asymmetry occurs
again but with the intensity of the SC peak at negative energy
becoming stronger. This feature has recently been confirmed by the
scanning tunneling microscopy (STM) experiments on
BaFe$_{2-x}$Co$_x$As$_2$~\cite{pan}.

In summary, we have examined theoretically the coexistence of the
SDW and SC orders in electron-doped iron-pnictide superconductors
based on the two orbital model and BdG equations. The phase diagram
is mapped out and the coexistence of the SDW and SC orders occurs at
low doping. The evolution of the Fermi surface as the doping varies
is presented and the results agree with several ARPES experiments.
The LDOS has also been calculated from low to high doping. The
signatures of the SDW order are identified and is consistent with
the recent STM experiment. It is important to point out that the
present results are critically dependent on the two orbital model
suggested in Ref.~\cite{zhang}.

We thank S. H. Pan and Ang Li for useful discussion and showing us
their STM data before publication. This work was supported by the Texas Center for
Superconductivity at the University of Houston and by the Robert A.
Welch Foundation under the Grant No. E-1146.

\end{document}